\documentclass[prl,twocolumn,superscriptaddress,showpacs]{revtex4}
\usepackage{amssymb,amsmath,graphicx,verbatim}

\newcommand{\ket}[1]{|#1\rangle}

\begin{document}
\title{Reply to ``Comment on `Role of Initial Entanglement and Non-Gaussianity in the Decoherence of Photon-Number Entangled States Evolving in a Noisy Channel' "}
\author{Michele Allegra}
\affiliation{ISI Foundation, I-10133 Torino, Italia.}
\affiliation{Dipartimento di Fisica Teorica, Universit\`a degli Studi di Torino, I-10125 Torino, Italia.}
\author{Paolo Giorda}
\affiliation{ISI Foundation, I-10133 Torino, Italia.}
\author{Matteo G. A. Paris}
\affiliation{Dipartimento di Fisica, Universit\`a degli Studi di Milano,
I-20133 Milano, Italia.}
\maketitle
In our Letter \cite{All10} we addressed the evolution of
photon-number entangled states $|\Psi\rangle = \sum_n \Psi_n
|n\rangle|n\rangle$ (PNES), $|n\rangle$ being a $n$-photon Fock state of
a harmonic oscillator, in noisy channels described by the master
equation $\dot\varrho= A \sum_{j=1}^2 \mathcal{L}[a_j] \varrho + B \sum_{j=1}^2 \mathcal{L}[a^\dag_j] \varrho$,
$\mathcal{L}[O]\varrho=2 O \varrho O^\dag - O^\dag O \varrho - \varrho O^\dag O$, $A=\frac12 \Gamma (1+N_T)$, $B=\frac12 \Gamma N_T$,
where $\Gamma$ is the damping factor of the channel and $N_T$ the
average number of thermal excitations of the channel. Upon
exploiting several non equivalent separability criteria we found evidence that
entanglement of Gaussian PNES survives longer, and thus
we drew a conjecture about the generality of this
result. General arguments supporting
the conjecture have been then put forward in \cite{Ade10}.
On the other hand, the conjecture needs not to be true
if one enlarges the set of separability criteria, e.g.
to include the negativity of the density matrix under
partial transposition (NDPT). 
Indeed, in \cite{Sab11} it has been shown that the conjecture
is too strong to be maintainable in the general case.
\par
The Comment to our Letter\cite{Comment} presents two examples where
the conjecture is violated. The first is
the states $\ket{\psi_{01}}=c_0 |00\rangle + c_1 |11\rangle$ with $0 \le c_1^2\le 1/2$,
whereas the second example
concerns photon subtracted squeezed vacuum (PSSV), i.e.
the PNES with $\psi_n=\propto (n+1) x^{n+1}$ with initial entanglement
$\epsilon_0\in\{0.1,1\}$ and initial energy $E_0\in\{0.013, 0.3\}$.
Not surprisingly, these examples confirm
that by enlarging the set of separability criteria
the conjecture is not maintainable.
However, these examples are worth to be analyzed in order
to assess whether their "robustness" compared to Gaussian states
has some relevant physical consequences.
To this aim we have evaluated the residual NDPT
$\mathcal{N}_R[\rho(t_G)]$ displayed by the above states
at the Simon' (Gaussian) separation time $t_G$ (i.e. at the time in which
Gaussian states with same initial entanglement/energy become separable)
for $B/A\equiv N_T/(1+N_T)\lesssim 0.5$
(remind that one photon at optical frequency $\nu\sim10^{15}$ Hz corresponds to
a temperature $T\sim 5\, 10^4$ K).
\par
\begin{figure}[h!]
\includegraphics[width=0.45\columnwidth]{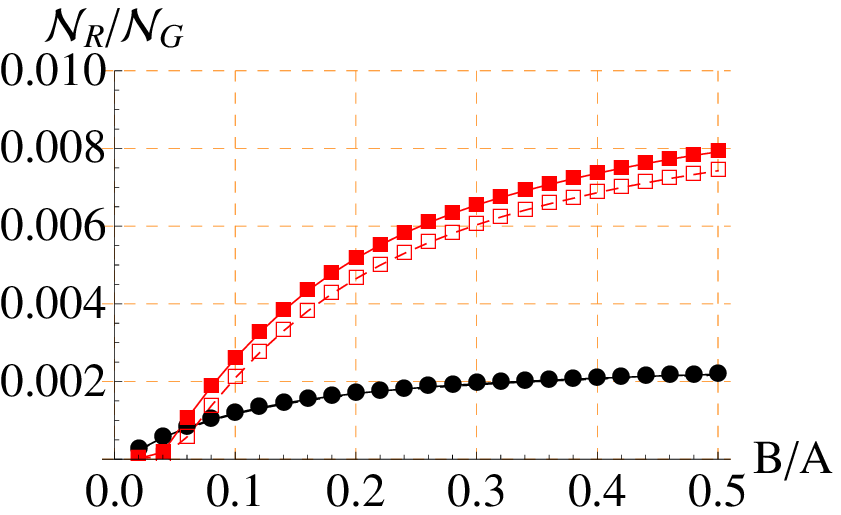}
\includegraphics[width=0.45\columnwidth]{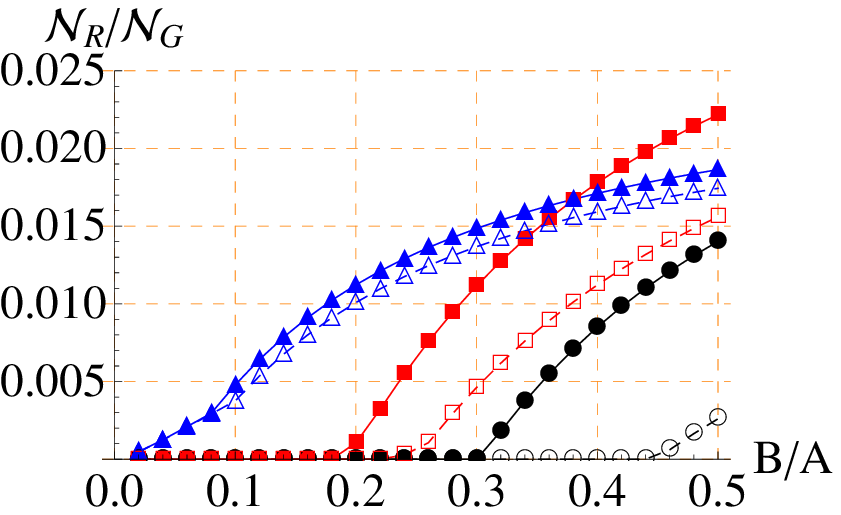}
\includegraphics[width=0.45\columnwidth]{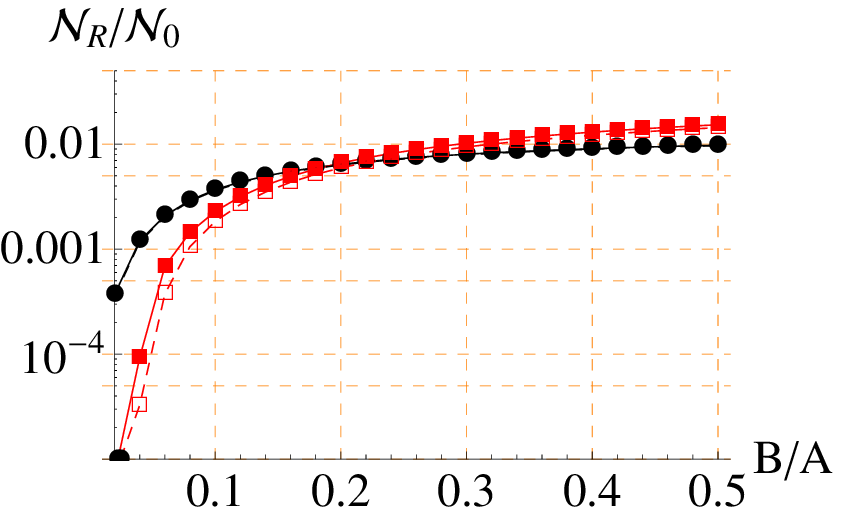}
\includegraphics[width=0.45\columnwidth]{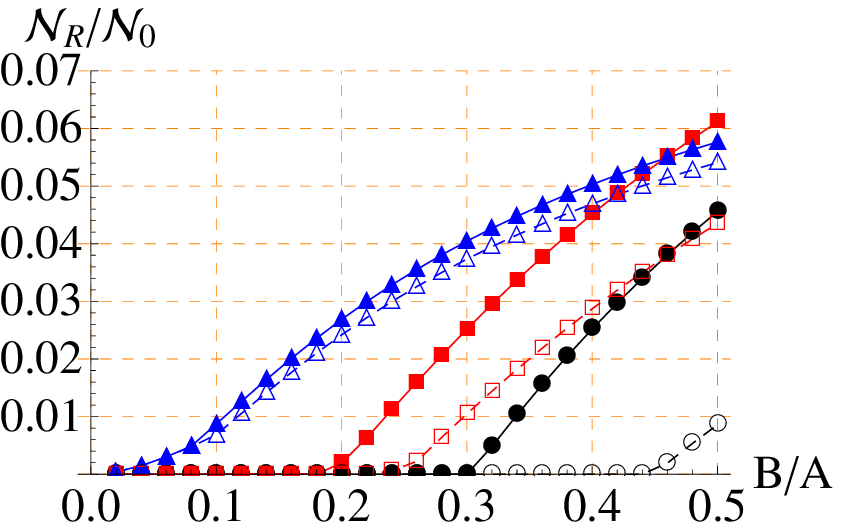}
\caption{
Renormalized negativity at separability time $t_{G}$ for Gaussian states
with same initial  energy (full symbols) or entanglement (open symbols). (Left)
PSSV states: black circles, $\epsilon_0 = 0.1, E_0=0.013$ (notice that the curves overlap); red squares, $\epsilon_0 = 1.0, E_0=0.3 $. (Right)
$\ket{\psi_{01}}$ states: black circles, $|c_1|^2 = 0.5$; red squares, $|c_1|^2 = 0.25 $;
blue triangles, $|c_1|^2 = 0.05 $.
\label{f:f1}}
\end{figure}

Results are shown in
Fig. \ref{f:f1}, where, in order to give an estimate of the level
of residual entanglement, we report the ratios
$\mathcal{N}_R/\mathcal{N}_G$ and
$\mathcal{N}_R/\mathcal{N}_0$, where
$\mathcal{N}_G$ is the maximal entanglement at the energy
of $\rho(t_G)$, i.e. the entanglement displayed by a pure Gaussian state with same energy,  and $\mathcal{N}_0$ is the initial
entanglement.
For PSSV states (left panels), we have that
although their separation time can be larger than $t_G$,  their residual
entanglement is very small for all values of $B/A$ and in particular
for realistic values of $N_T$, i.e. $B/A \ll 1$.  Similar
results are obtained for $\ket{\psi_{01}}$
(right panels). 
\par
In summary, the conjecture that we put forward in our Letter, based on a correct comparison
of several different separabililty criteria, is not maintanable in a strict sense if one includes the NDPT criterion.
Nevertheless, for the examples of non Gaussian states proposed in the Comment, which violate the conjecture,
the residual entanglement is extremely low. Therefore it remains an open question whether these, 
as well as other kinds of states, could represent a useful resource in quantum communication protocols.
The issue of degradation of continuous variables entanglement in noisy channels is indeed a
extremely relevant one, and it is worth be to analyzed in detail with a more
substantial analysis.

\end{document}